\documentclass[prl,twocolumn,aps,showpacs,amsmath,amssymb]{revtex4}
\usepackage{graphicx}
\usepackage{dcolumn}
\usepackage{bm}

\begin{document}

\title{Role of the Majorana Fermion and the Edge Mode in Chiral Superfluidity near a $p$-Wave Feshbach Resonance
}

\author{T. Mizushima}
\affiliation{Department of Physics, Okayama University,
Okayama 700-8530, Japan}
\author{M. Ichioka}
\affiliation{Department of Physics, Okayama University,
Okayama 700-8530, Japan}
\author{K. Machida}
\affiliation{Department of Physics, Okayama University,
Okayama 700-8530, Japan}
\date{\today}

\begin{abstract}

The visualization of chiral $p$-wave superfluidity in Fermi gases near $p$-wave Feshbach resonances is theoretically examined. It is proposed that the superfluidity becomes detectable in the entire BCS-BEC regimes through (i) vortex visualization by the density depletion inside the vortex core and (ii) intrinsic angular momentum in vortex free states. It is revealed that both (i) and (ii) are closely connected with the Majorana zero energy mode of the vortex core and the edge mode, which survive until the strong coupling BCS regime is approached from the weak coupling limit and vanish in the BEC regime.

\end{abstract}



\maketitle

There has been intense interest on $p$-wave pairing superfluids recently in ultracold atom systems both experimentally and theoretically. It is highly expected that $p$-wave superfluidity may be realized experimentally, by using $p$-wave Feshbach resonance \cite{p-wave} which is now found for several atomic species $^6$Li and $^{40}$K under magnetic field sweeping. In the two-dimensional (2D) system with the presence of magnetic field, the superfluid has a spinless chiral $p$-wave symmetry \cite{botelho,ptheory}. A theoretical task to be urgently clarified is how to detect the superfluidity. In the $s$-wave pairing case, the vortex imaging through the density depletion at a vortex core is decisive to establish it \cite{zwierlein}. There, the density depletion at vortex core is caused by the spontaneous particle-hole asymmetry \cite{hayashi} of the discrete core-localized states, {\it i.e.}, so-called Caroli-de Gennes-Matricon (CdGM) states \cite{cdgm}. It is not quite obvious that the same is true for $p$-wave pairing \cite{kopnin,ivanov,matsumoto}, whose detailed study is one of the purposes here. 

The CdGM states in chiral $p$-wave superfluid are intriguing on its own right because it has exact zero energy state (ZES) as shown in the BCS limit where the chemical potential $\mu$ is equal to the Fermi energy $E_F$~\cite{kopnin,ivanov,matsumoto}. Since the ZES can be described by the one-dimensional (1D) Majorana equation \cite{tewari}, the quasiparticle of the ZES is a Majorana fermion. Different from ZES of the Jackiw-Rebbi solution in the 1D Dirac equation at the domain wall \cite{jackiw,ssh,machida}, the remarkable feature of Majorana fermions is that their creation is expressed by the self-Hermitian operator, and their host vortices obey the non-abelian statistics \cite{read,ivanov}. These can be utilized for quantum computer \cite{kitaev}. It is natural to expect that the observation of $p$-wave Feshbach resonance \cite {p-wave} opens a door connecting ultracold atoms to diverse research fields, such as quantum computation, the condensed matter, and the relativistic field theory.

The other low-lying mode characteristic to $p$-wave superfluid is the edge mode, resulting from the Andreev scattering at the rigid wall of the system \cite{stone2}. Due to the edge mode, the $p$-wave superfluid has the spontaneous mass flow in the vortex free state, which accompanies the angular momentum arising from the internal motion of $p$-wave pairs \cite{kita}. The observation of macroscopic angular momentum could be used as another method to detect evidence of the $p$-wave superfluidity.

The merit of study in atomic gases is that we can manipulate the interaction through Feshbach resonance. The Cooper pairs realized within $\mu \!>\! 0$ are turned to strong coupling at $\mu \rightarrow +0$, and further to Bose-Einstein condensation (BEC) of molecular bosons when $\mu \!<\! 0$. Contrasted to the BCS-BEC crossover in the s-wave pairing, there is a topological phase transition at $\mu \!=\! 0$ between BCS and BEC in the $p$-wave pairing~\cite{read}. The purpose of this Letter is to examine the possibility to experimentally detect the evidence of superfluidity associated with $p$-wave Feshbach resonances, in the entire range from BCS to BEC regimes. It is proposed that the superfluidity becomes detectable through the vortex visualization and intrinsic angular momentum. From fully microscopic point of view based on the Bogoliubov-de Gennes (BdG) equation, which enables the systematic study in the BCS-BEC regimes, we reveal that their visualization is closely connected with novel quasiparticle states, such as the Majorana ZES and the edge mode.

The attractive $p$-wave interaction by the Feshbach resonance can produce a pairing between spinless fermions \cite{botelho,ptheory}. 
The pair potential for chiral $p$-wave pairing in 2D systems can be expressed as
$
\Delta ({\bm r},{\bm k}) \!=\! - \frac{k_x +i k_y}{\sqrt{2}k_F} \Delta _{+1} ({\bm r})
+ \frac{k_x-i k_y}{\sqrt{2}k_F} \Delta _{-1} ({\bm r})
$,
in terms of the eigenstate of the orbital angular momentum $\hat{L}_z|m\rangle \!=\! m| m\rangle$ ($m\!=\! 0, \pm 1$), 
where ${\bm r}$ and ${\bm k}$ are the center-of-mass coordinate and the relative wave vector of the pair. 
Under the pair potential $\Delta({\bm r},{\bm k})$, the quasiparticle eigenstates with the wave function $[u_{\bm q}, v_{\bm q}]^T$ is described by the BdG equation \cite{matsumoto, mizushima08}, 
\begin{eqnarray}
\left[
\begin{array}{cc}
H_0({\bm r}) & \Pi({\bm r}) \\ - \Pi^{\ast}({\bm r}) & -H_0({\bm r})
\end{array}
\right]
\left[\begin{array}{c}
u_{\bm q}({\bm r}) \\ v_{\bm q}({\bm r}) 
\end{array}\right]
= E_{\bm q}
\left[\begin{array}{c}
u_{\bm q}({\bm r}) \\ v_{\bm q}({\bm r})
\end{array}\right],
\label{eq:bdg}
\end{eqnarray}
where $\hbar\!=\! k_B \!=\! 1$. The applicability of the single channel model in the $p$-wave case is discussed in Ref.~\cite{gurarierAP}, where it is verified that this simple theory is suitable for describing the BCS-to-BEC regimes associated with narrow $p$-wave Feshbach resonances. 
We set $\Pi ({\bm r}) \!=\! - \frac{i}{k_F} \sum _{\pm}\{ 
\Delta _{\pm 1}({\bm r}) \square _{\pm} +\frac{1}{2}\square _{\pm}\Delta _{\pm 1}({\bm r})
\}$ and $H_0({\bm r}) \!=\! -\frac{\nabla^2}{2M}+V({\bm r})-\mu$ with mass $M$, $k_F \!=\! \sqrt{2ME_F}$, and the harmonic trap potential $V({\bm r})$ which confines atomic gases. While we mainly discuss the case when $V({\bm r}) \!=\! 0$, we also confirmed that the presence of a harmonic trap does not change essential features obtained with $V\!=\!0$. 
In the 2D polar coordinates ${\bm r} \!=\! (r,\theta)$, $\square _{\pm} \!=\! \mp \frac{e^{\pm i \theta}}{\sqrt{2}}(\partial _r \pm \frac{i}{r}\partial _{\theta})$. Assuming the cylindrical symmetry of the pair potential, $\Delta _m({\bm r}) \!=\! \Delta _m(r) e^{iw_{m} \theta}$ with the winding number $w_{m}\!\in\!\mathbb{Z}$. Thus, the eigenstates labeled by ${\bm q}$ in Eq.~(\ref{eq:bdg}) are characterized by the azimuthal quantum number $q_{\theta} \!\in\! {\mathbb Z}$  
as $u_{\bm q}({\bm r})\!=\!u_{\bm q}(r) e^{iq_{\theta}\theta}$, $v_{\bm q}({\bm r})\!=\!v_{\bm q}(r) e^{i(q_{\theta}-w_{+1}-1)\theta}$, and we find $w_{-1} \!=\! w_{+1} + 2$. In this work, we consider the case when the dominant chiral component is $\Delta_{+1}({\bm r})$ and has vortex winding $w_{+1} \!=\!-1$, {\it i.e.}, vortex current flows along the direction opposite to the orbital motion of the chiral pairing. The passive component $\Delta_{-1}({\bm r})$ are only induced around the vortex core and the boundary. The other cases of vortex winding will be discussed in Ref.~\cite{mizushima08}.

The pair potentials $\Delta _{\pm 1}$ are determined by the gap equation
$\Delta _{\pm 1} ({\bm r}) \!=\! \frac{ig}{k _F} \sum _{E_{\bm q}<0}
[ v^{\ast}_{\bm q}({\bm r}) \square _{\mp} u_{\bm q}({\bm r}) 
- u_{\bm q}({\bm r})\square _{\mp} v^{\ast}_{\bm q}({\bm r}) ]$ at zero temperature ($T\!=\! 0$) \cite{matsumoto}.
From practical point of view, it is convenient to express the coupling constant $g$ with the two-body bound state energy in vacuum $E_b$, as $g^{-1} \!=\! - \int\frac{d{\bm k}}{(2\pi)^2} \frac{(k/k_F)^2}{k^2/M - E_b}$ \cite{randeria,botelho}. Note that in contrast to the $s$-wave case, the resulting gap equation still contains the logarithmic divergence on the energy cutoff $E_c$ \cite{randeria}. Throughout this work, we calculate the set of equations with $E_c \!=\! 20$, $40$, and $60E_F$ and verify that there is no qualitative difference between them. Hence, in this Letter, we present the results with $E_c\!=\! 60E_F$. We solve the set of equations in a cylinder with the radius $R\!=\! 50k^{-1}_F$ and the boundary condition $u_{\bm q}(r\!=\! R) \!=\! v_{\bm q}(r\!=\! R)\!=\! 0$. Here, all the quantities are scaled by the energy unit $E_F$ and the length unit $k_F^{-1}$. During the self-consistent calculation, we vary $\mu$ so as to fix the total particle number $N \!=\! \int \rho({\bm r})d{\bm r} \!=\! 604$, with  the particle density 
$\rho({\bm r}) \!=\! \sum _{E_{\bm q}<0}|u_{\bm q}({\bm r})|^2$. 
When changing the pairing interaction by varying $E_b$ within $-1.8\!\le\!E_b/E_F\!<\!1.0$, the chemical potential and the maximum of the pair potential $\Delta \!\equiv\! \max|\Delta _{+1}(r)|$ continuously change from $\mu \!=\! 0.6E_F$ and $\Delta \!=\! 0.79E_F$ at $E_b\!=\! 1.0E_F$  of strong coupling BCS to $\mu\!=\!-0.21E_F$ and $\Delta \!=\! 1.46E_F$ at $E_b\!=\!-1.8E_F$ of BEC.


\begin{figure}[t!]
\includegraphics[width=0.8\linewidth]{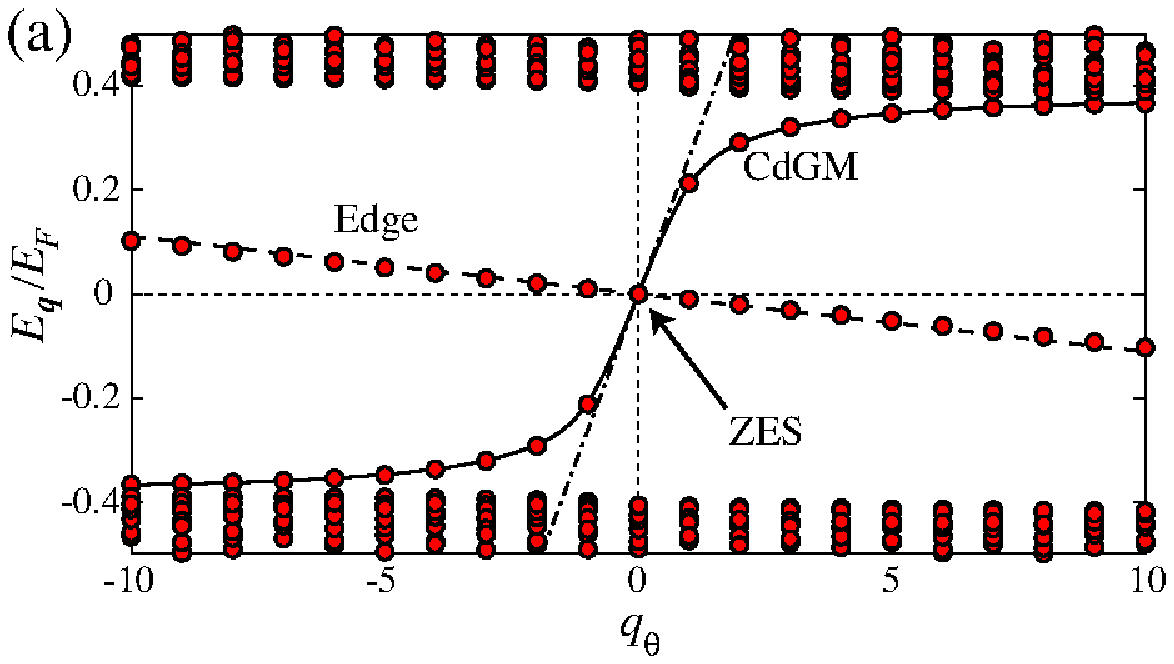}
\includegraphics[width=0.85\linewidth]{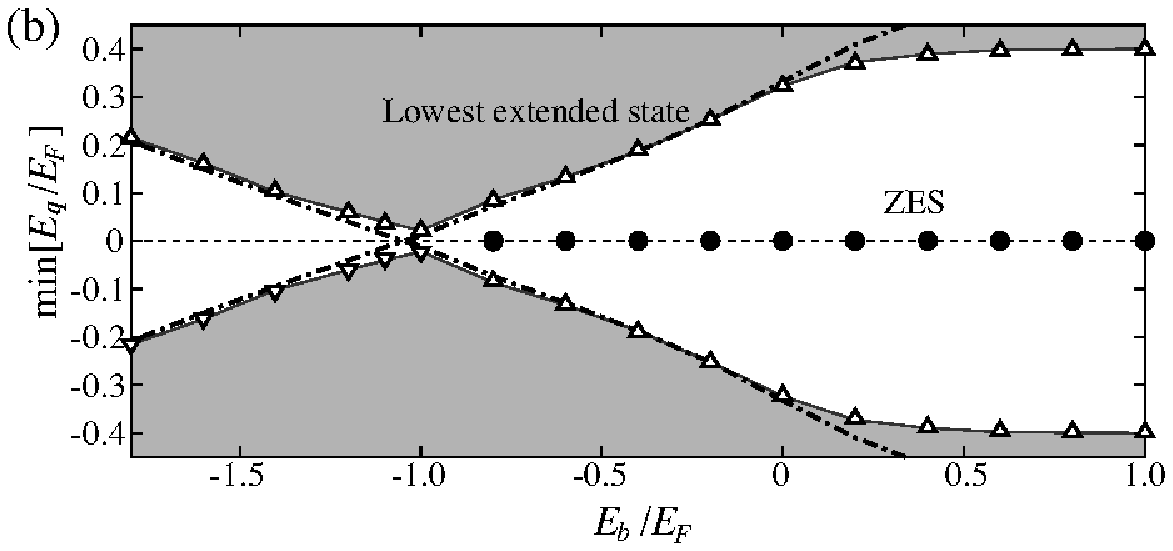}
\caption{(Color online) 
(a) 
Energy spectrum $E_{\bm q}$ of the $w_{+1} \!=\! -1$ vortex as functions of $q_{\theta}$, at $E_b\!=\! 1.0E_F$ with $\mu \!=\! 0.6E_F$ and $\Delta \!=\! 0.79E_F$. 
The dashed and dashed-dotted lines denote $E^{({\rm e})}_{\bm q}$ and $E^{({\rm c})}_{\bm q}$. 
(b) 
Low-lying eigenenergies at $q_{\theta} \!=\! 0$ in the $w_{+1} \!=\! -1$ vortex as functions of $E_b$. 
Circles and triangles denote the eigenenergies of the ZES and the lowest bulk excitation state, respectively. 
The dotted-dashed line denotes $\pm |\mu|$. 
}
\label{fig:gap}
\end{figure}

In Fig.~\ref{fig:gap}(a), we plot the energy spectrum of the vortex state at $E_b\!=\! 1.0E_F$. There is a symmetry $E_{q_{\theta}} \!=\! -E_{-q_{\theta}}$. Two branches appear inside the bulk energy gap $E_{\rm gap}=\!0.4E_F$. The branch labeled as ``Edge'' in Fig.~\ref{fig:gap}(a) consists of the eigenstates with a wave function localized at the edge of the cylinder as the Andreev bound state. The other branch labeled as ``CdGM'' is composed by the core-localized mode. 
The energy separations between discrete levels of the CdGM states are large in this strong coupling. In the BCS limit, dispersion relations of the branches can be analytically derived from Eq.~(\ref{eq:bdg}) as $E^{({\rm e})}_{\bm q} \!=\! - ( q_{\theta}-\frac{w_{+1}+1}{2}) \frac{\Delta/\sqrt{2}}{k_FR}$ for edge mode by Stone and Roy~\cite{stone2}, and 
$
E^{({\rm c})}_{\bm q} \!=\! - ( q_{\theta}-\frac{w_{+1}+1}{2}) \omega _0
+ ( n - \frac{w_{+1}+1}{2} ) \omega _1 $  
with $n \!\in\! {\mathbb Z}$, $w_{+1} \!\neq\! 0$ and $\omega _0 \!\sim\! \Delta^2/E_F \!\ll\! \omega _1 \!\sim\! \Delta$ for CdGM mode using $|q_{\theta}|\!\ll\! k_F\xi$ and $\Delta \!\ll\! \mu \!\sim\! E_F$ by us~\cite{mizushima08}.  
These analytical expressions in the BCS limit, presented by lines in Fig. ~\ref{fig:gap}(a), still give nice fitting to our numerical results of the strong coupling. It is noted that the eigenenergies of $E^{({\rm e})}_{\bm q}$ and $E^{({\rm c})}_{\bm q}$ can become exactly zero in vortices with arbitrary odd number of $w_{+1}$.

We notice that the behaviors of the $E_b$-dependence of ZES and the density depletion can be classified to three regimes: (i) the BCS limit of weak coupling with $\mu \!\simeq\! E_F$ ($E_b/E_F\!\gg\! 1$), (ii) the resonance regime of strong coupling BCS with $0\!<\!\mu\!\ll\!E_F$ ($|E_b|/E_F \!<\! 1.0$), (iii) the BEC regime with $\mu \!<\! 0$ ($E_b \!<\! -1.0E_F$). The topological phase transition occurs at $\mu = 0$ ($E_b = -1.0E_F$). 

The low-lying eigenenergies of the spectrum with $q_{\theta} \!=\! 0$ are displayed in Fig.~\ref{fig:gap}(b) as a function of $E_b$. The lowest extended states denoted by triangle points correspond to the bulk excitation gap $E_{\rm gap}$. In the resonance regime (ii), $E_{\rm gap} \sim \mu$ and $E_{\rm gap}\rightarrow 0$ approaching the topological phase transition in the $p$-wave pairing. The bulk excitation in the BEC regime (iii) is again gapful as $E_{\rm gap} \sim -\mu$, corresponding to the binding energy of molecules. Within the excitation gap, doubly degenerate ZES's of CdGM and edge modes behave as follows. In the BCS limit, it is known from $E^{({\rm c})}_{\bm q}$ and the index theorem \cite{volovik99,tewari} that the ZES always appears. Figure~\ref{fig:gap}(b) shows that the ZES survives also in the resonance regime beyond the BCS limit, until just before the topological phase transition at $E_b \!=\! -1.0E_F$. We have also confirmed the degenerate ZES's appear in harmonically trapped systems \cite{mizushima08}. In the BEC regime beyond the topological transition, the ZES vanishes and merges to the excitation gap. In the vicinity of $E_b\!=\!-1.0E_F$, the ZES shifts to finite energies by the interference of the CdGM and the edge states. The interference of the ZES's becomes effective when $\mu \rightarrow +0$, because the length scale of the spatial variation of $u_{\bm q}$ and $v_{\bm q}$ becomes comparable with the system size $R$ ~\cite{gurarie}. This shift of ZES is seen at $E_b\!=\! -1.0E_F$ in Fig.~\ref{fig:gap}(b), since this change occurs slightly before $E_b\!=\! -1.0E_F$ when the system radius $R$ is finite.

\begin{figure}[b!]
\includegraphics[width=0.85\linewidth]{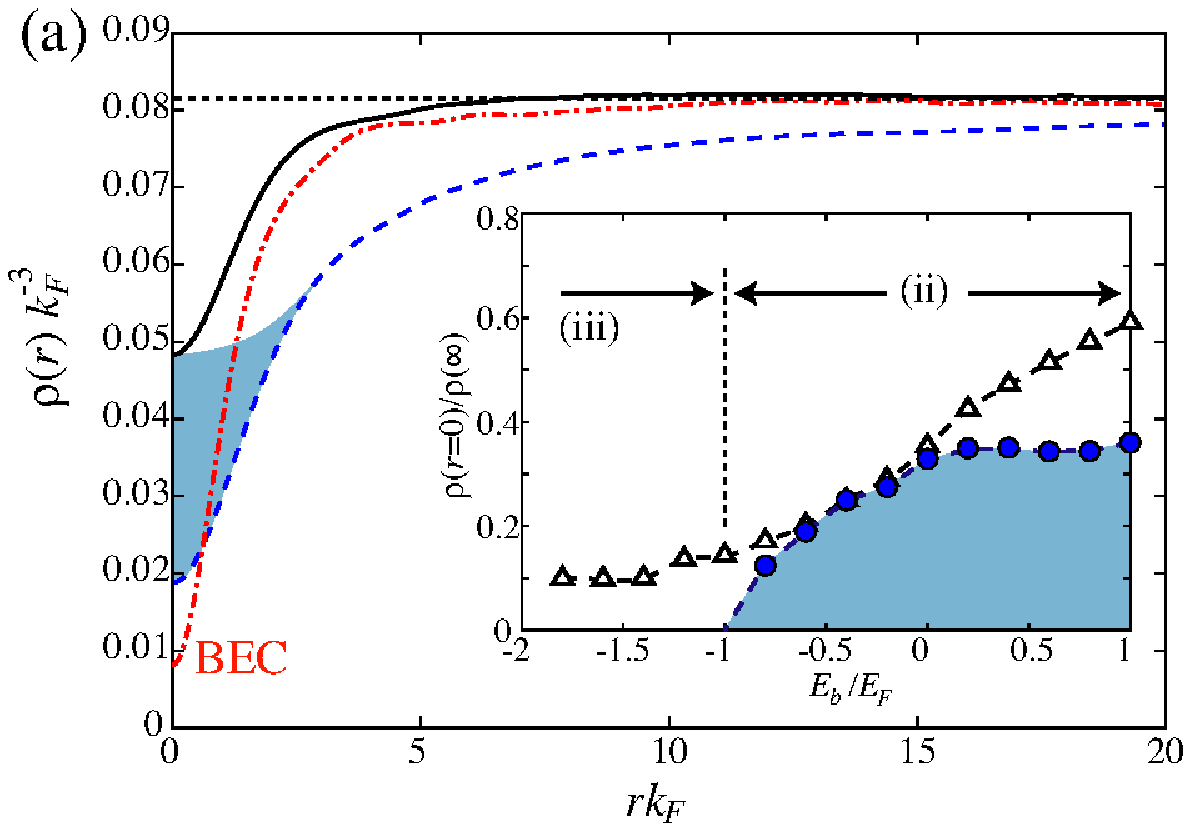}
\includegraphics[width=0.85\linewidth]{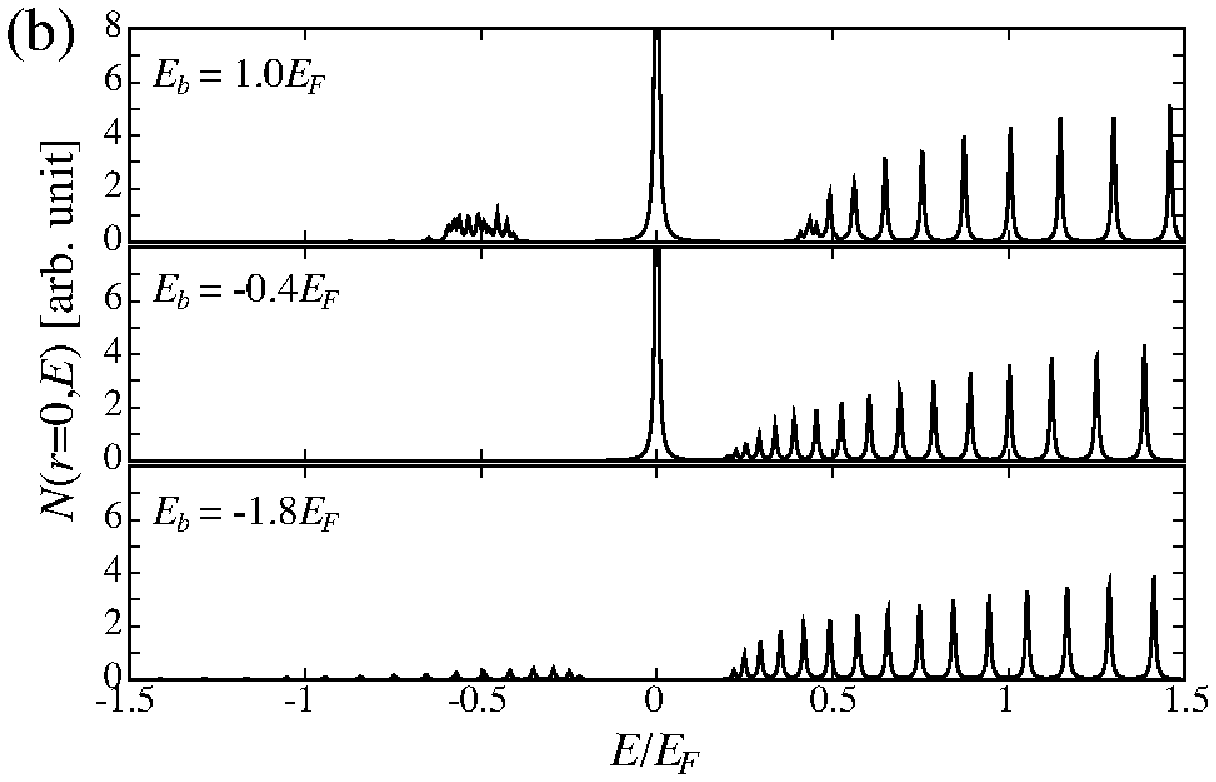}
\caption{(Color online) 
(a) 
Particle density $\rho(r)$ in the $w_{+1} \!=\! -1$ vortex state at $E_b \!=\! 1.0E_F$ (solid line) and the BEC regime $E_b \!=\! -1.8E_F$ (dotted-dashed line). The dashed line shows the contribution of the extended states at $|E_{\bm q}| \!>\! 0.4E_F$ to $\rho(r)$ at $E_b \!=\! 1.0E_F$, and the shaded area depicts the contribution of the ZES. The dotted line denotes constant $\rho(r)$ in the BCS limit \cite{matsumoto}. The inset in (a) shows the density depletion at the core $\rho(r\!=\!0)/\rho(\infty)$ (triangles) and the contribution of the ZES (circles). (b) Local density of states $\mathcal{N}(r\!=\! 0, E)$ at $E_b/E_F \!=\! 1.0$, $-0.4$, and $-1.8$.
}
\label{fig:dns}
\end{figure}

The quantized vortex which is a hallmark of superfluidity is detected by the depletion of the particle density $\rho(r)$ at the vortex core. Here, we discuss the relation of the ZES and vortex visualization by $\rho(r\!=\!0)$ at the core. To see the contribution of the ZES and the extended states outside the gap, in Fig.~\ref{fig:dns}(b) we present the local density of states (LDOS) at $r\!=\!0$ defined as $\mathcal{N}(r,E) \!=\! \sum _{\bm q}|u_{\bm q}({\bm r})|^2\delta(E-E_{\bm q})$. In the BCS limit, it is known that $\rho(r)$ keeps constant even inside the vortex core when $w_{+1}\!=\! -1$~\cite{matsumoto}, as presented by dotted line in Fig.~\ref{fig:dns}(a). This indicates that the vortex is completely invisible through the density imaging in the BCS limit. As approaching the resonance regime from the BCS limit, the vortex core gradually becomes visible by the contribution of extended states. We show $\rho(r)$ at $E_b\!=\! 1.0E_F$ by solid line in Fig.~\ref{fig:dns}(a). If the contributions of CdGM states are removed from $\rho(r)$, the density depletion by the extended states is much deeper as shown by dashed line in Fig.~\ref{fig:dns}(a), which reflects the spatial structure of $\Delta _{+1}(r)$. This strong depletion at the vortex core is compensated by the ZES contribution, as shown by the shade in  Fig.~\ref{fig:dns}(a), and it makes the depletion weaker. This contribution of ZES is absent in the case of $s$-wave pairing. Thus, the density contrast of vortex core visualization is weaker in the chiral $p$-wave pairing. The vortex visibility $\rho(r\!=\!0)/\rho(r\!=\!\infty)$ and the contribution of the ZES are plotted in the inset of Fig.~\ref{fig:dns}(a) as a function of $E_b$. When $-1 \!<\! E_b/E_F \!<\! 0$ in the resonance regime (ii), $\rho(r\!=\!0)$ comes from only the ZES, since the LDOS at $E_b\!=\! -0.4E_F$ in Fig.~\ref{fig:dns}(b) has no intensity outside the gap for $E_{\bm q}\!<\!0$. Note that the absence of the negative energy band results from the strong coupling effects, such as $|\mu| \!\ll\! \Delta\!\sim \!E_F$, while there still remains the particle-hole symmetry. In the BEC regime, the vortex core is clearly visible through the density imaging experiments~\cite{zwierlein}. As shown in Fig.~\ref{fig:dns}(b) for $E_b\!=\! -1.8E_F$, the ZES disappears and the excitation spectrum becomes gapful. Thus, as the dotted-dashed line in Fig.~\ref{fig:dns}(a), $\rho(r)$ approaches perfect depletion  $\rho(r) \propto |\Delta _{+1}(r)|^2$ in the BEC. 

\begin{figure}[t!]
\includegraphics[width=\linewidth]{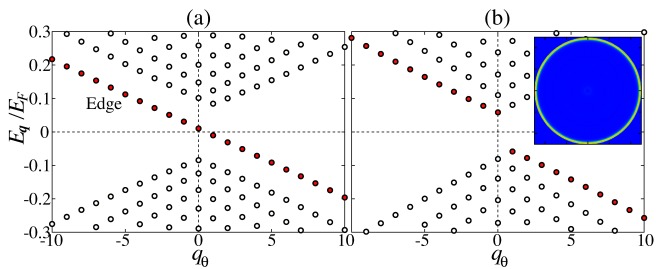}
\caption{(Color online) 
Energy spectrum $E_{\bm q}$ of the vortex free state near the resonance: (a) $E_b\!=\! -0.8E_F$ with $\mu \!=\! 0.07E_F$ and $\Delta/E_F \!=\! 1.29$ and (b) $E_b\!=\! -1.2E_F$ with $\mu \!=\! -0.05E_F$ and $\Delta/E_F \!=\! 1.36$. 
The inset in (b) depicts the current density $|j_{\theta}(r)|$ at $E_b\!=\!-1.2E_F$ within $|x|, |y|\!<\! 50k^{-1}_F$.
}
\label{fig:egnsSC}
\end{figure}

Next, we discuss the relation of energy spectrum and intrinsic angular momentum in the vortex free state. In Fig.~\ref{fig:egnsSC}(a) we plot the energy spectra of the vortex free state with $w_{+1} \!=\! 0$ at $E_b\!=\! -0.8E_F$ in the resonance regime. There we see the linear branch of the edge mode at low energies within the gap. In the vortex free state, the edge mode has small but finite energy at $q_\theta=0$, different from the vortex states. The edge states with negative energies are occupied at $T\!=\! 0$. Due to asymmetry on $q_{\theta}$, the occupied edge modes with $q_{\theta}>0$ produce the spontaneous mass current \cite{stone2} $j_{\theta}(r) \!\equiv\! \sum _{E_{\bm q}\!<\! 0} \frac{q_{\theta}}{r}|u_{\bm q}(r)|^2$ around the edge of cylinder. Hence, net angular momentum $L_z \!=\! \int r j_{\theta}(r)d{\bm r}$ is not zero even in the vortex free state. We numerically find that $L_z/N\!=\! 0.509$ at $E_b \!=\! -0.8E_F$. This is also understandable as the angular momentum carried by the orbital motion of the Cooper pair with the chirality $m\!=\!+1$. The small deviation from the intuitive expectation $+\frac{1}{2}$ is due to the pairing $\Delta _{-1}(r)$ with $w_{-1}\!=\!+2$ induced around the edge.

More noteworthy is that there still exists the spontaneous mass flow even in the BEC regime with $\mu \!<\! 0$. 
Figure~\ref{fig:egnsSC}(b) shows that the branch of the edge mode with the positive (negative) eigenenergy merges into the upper (lower) band of the bulk excitations, and the resulting energy spectrum becomes gapful with the gap $\pm|\mu|$. Hence, the whole spectrum remains asymmetry with respect to $q_{\theta}$, which can produce the spontaneous mass flow along the edge, even without edge modes in the BEC regime. The inset in Fig.~\ref{fig:egnsSC}(b) shows $j_{\theta}(r)$, which is localized at the edge within $\xi \!\sim\! k^{-1}_F$. It is also found that $L_z/N$ stays around $0.5$ within $15\%$ in the range $-1.8\!\le\! E_b/E_F \!\le\! 1.0$.

\begin{figure}[b!]
\includegraphics[width=\linewidth]{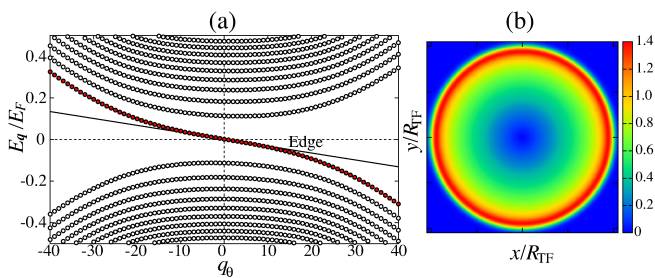}
\caption{(Color online) 
(a) Energy spectrum $E_{\bm q}$ of the vortex free state with $w_{+1} \!=\! 0$ in harmonically trapped state, where $\mu\!=\!0.78E_F$ and $\max{|\Delta _{+1}(r)|} \!=\! 1.05E_F$ with $E_F /\omega \!=\! 42.5$ at $E_b\!=\!0.0$. 
The solid line denotes $E^{({\rm e})}_{\bm q}$ with $R\!\equiv\!R_{\rm TF}\!=\! 74.5k^{-1}_F$ and $\Delta \!\equiv\! |\Delta _{+1}(r\!=\!0.85R_{\rm TF})| \!=\! 0.35E_F$. (b) The corresponding current density $|j_{\theta}(r)|$ within $|x|, |y|\!<\! 1.05R_{\rm TF}$.
}
\label{fig:trap}
\end{figure}

Finally, we examine influences of a trap potential $V({\bm r})$ on the edge mode and the mass flow. 
Figure~\ref{fig:trap}(a) depicts the energy spectrum of $N\!=\!1,000$ atoms in the vortex free state under a harmonic trap $V({\bm r}) \!=\! \frac{1}{2}m\omega^2r^2$, which still accompanies low-lying edge modes. The presence of the edge mode in a harmonic trap was discussed in Ref.~\cite{stone3} under a given uniform pairing field $\Delta _{+1}(r)\!=\! {\rm const}$. We examine the edge mode by the full self-consistent calculation including spatial variation of $\Delta _{\pm 1}(r)$. In the presence of $V(r)$, since the wave functions of the edge mode have a broad peak around $r \!\sim\! 0.85R_{\rm TF}$ with Thomas-Fermi radius $R_{\rm TF}$, the spontaneous current $j_{\theta}(r)$ has broad profile around $R_{\rm TF}$, as shown in Fig.~\ref{fig:trap}(b). However, this does not seriously change the result  $L_z/N\!\sim\! 0.5$. 

The azimuthal flow along the Thomas-Fermi edge may affect the collective surface excitations, such as the quadrupole mode, which gives rise to the frequency shift between the co- and counter-rotating surface motion \cite{madison}. 
Hence, the measurement of the intrinsic angular momentum $L_z/N\!\simeq\! 0.5$, through the shift of the quadrupole frequencies, provides evidence of $p$-wave superfluidity.

In summary, we have theoretically investigated novel features of chiral superfluid Fermi gases in the strong coupling BCS regime and the BEC regime near $p$-wave Feshbach resonance. Majorana zero energy states of the vortex core mode and the edge modes survives from the BCS limit ($\mu \!\simeq\!E_F$) to the BCS-BEC transition point ($\mu \!=\!0$), and vanish in the BEC regime. In the relation to the low-energy modes, we estimate (i) vortex visualization by the density depletion and (ii) the intrinsic angular momentum in the vortex free state, 
finding that these can be used to experimentally detect the appearance of superfluidity in wide range from BCS to BEC through the $p$-wave Feshbach resonance. 




\begin{thebibliography}{99}

\bibitem{p-wave}
Y. Inada {\it et al.}, arXiv:0803.1405 and references therein.


\bibitem{ptheory}
T.-L. Ho and R.B. Diener, Phys. Rev. Lett. {\bf 94}, 090402 (2005); 
Y. Ohashi, {\it ibid} {\bf 94}, 050403 (2005); 
V. Gurarie, L. Radzihovsky, and A.V. Andreev, {\it ibid} {\bf 94}, 230403 (2005); 
C.-H. Cheng and S.-K. Yip, {\it ibid} {\bf 95}, 070404 (2005); 
M. Iskin and C.A.R. S\'{a} de Melo, {\it ibid} {\bf 96}, 040402 (2006).

\bibitem{botelho}
S. S. Botelho and C.A.R. S\'{a} de Melo, J. Low Temp. Phys. {\bf 140}, 409 (2005).

\bibitem{zwierlein}
M.W. Zwierlein {\it et al.}, Nature (London) {\bf 435}, 1047 (2005).


\bibitem{hayashi}
N. Hayashi {\it et al.}, J. Phys. Soc. Jpn. {\bf 67}, 3368 (1998); 
Phys. Rev. Lett. {\bf 80}, 2921 (1998).


\bibitem{cdgm}
C. Caroli, P.G. de Gennes, and J. Matricon, Phys. Lett. {\bf 9}, 307 (1964).

\bibitem{kopnin}
N.B. Kopnin and M.M. Salomaa, Phys. Rev. B {\bf 44}, 9667 (1991).

\bibitem{ivanov}
D. A. Ivanov, Phys. Rev. Lett. {\bf 86}, 268 (2001). 

\bibitem{matsumoto}
M. Matsumoto and R. Heeb, Phys. Rev. B {\bf 65}, 014504 (2001).

\bibitem{tewari}
S. Tewari, S. Das Sarma, and D.-H. Lee, Phys. Rev. Lett. {\bf 99}, 037001 (2007). 

\bibitem{jackiw}
R. Jackiw and C. Rebbi, Phys. Rev. D {\bf 13}, 3398 (1976).

\bibitem{ssh}
R. Jackiw and J.R. Schrieffer, Nucl. Phys. {\bf B190}, 254 (1981).

\bibitem{machida}
K. Machida and H. Nakanishi, Phys. Rev. B {\bf 30} 122 (1984); 
K. Machida and M. Fujita, Phys. Rev. B {\bf 30}, 5284 (1984).

\bibitem{read}
N. Read and D. Green, Phys. Rev. B {\bf 61}, 10267 (2000).

\bibitem{kitaev}
A. Kitaev, Ann. Phys. (N.Y.) {\bf 303}, 2 (2003).




\bibitem{stone2}
M. Stone and R. Roy, Phys. Rev. B {\bf 69}, 184511 (2004).

\bibitem{kita}
See for example, T. Kita, J. Phys. Soc. Jpn. {\bf 65}, 664 (1996).

\bibitem{mizushima08}
T. Mizushima {\it et al.}, 
in preparation.

\bibitem{gurarierAP}
V. Gurarie and L. Radzihovsky, Ann. Phys. {\bf 322}, 2 (2007).

\bibitem{randeria}
M. Randeria, J.-M. Duan, and L.-Y. Shieh, 
Phys. Rev. B {\bf 41}, 327 (1990).



\bibitem{volovik99}
G.E. Volovik, Pis'ma Zh. Eksp. Teor. Fiz. {\bf 70}, 601 (1999) [JETP Lett. {\bf 70}, 609 (1999)].






\bibitem{gurarie}
V. Gurarie and L. Radzihovsky, Phys. Rev. B {\bf 75}, 212509 (2007).





\bibitem{stone3}
M. Stone and I. Anduaga, arXiv:cond-mat/0702412.


\bibitem{madison}
F. Chevy, K.W. Madison, and J. Dalibard, Phys. Rev. Lett. {\bf 85}, 2223 (2000).

\end{thebibliography}
\end{document}